\documentclass[twocolumn,showpacs,preprintnumbers,amsmath,amssymb,pre]{revtex4}

\usepackage{graphicx}
\usepackage{dcolumn}
\usepackage{bm}

\begin{document}


\title{Market dynamics after large  financial crash}

\author{G. Buchbinder}
\email{glb@omsu.ru}
\author{K. Chistilin}%
 \email{Chistilin_K@mail.ru}
\affiliation{
Physics Department, Omsk State University, Peace Avenue, 55a, 644077 Omsk, Russia\\
}%


\date{\today}

\begin{abstract}
The model describing  market dynamics after a large financial crash is considered in terms of the stochastic differential equation
 of Ito. Physically, the model presents an overdamped Brownian particle  moving in the nonstationary one-dimensional potential $U$
 under  the influence of the variable noise intensity, depending on the particle position $x$.
 Based on the empirical data the approximate estimation of the Kramers-Moyal coefficients $D_{1,2}$ allow to predicate quite definitely the behavior of
 the potential introduced by $D_1 = - \partial U /\partial x$ and the volatility $\sim \sqrt{D_2}$. It has been shown that the presented model describes  well enough the best known empirical facts relative to the large financial crash of October 1987.
\end{abstract}

\maketitle

\section{\label{sec:Intro}Introduction}
The dynamics of the financial markets has been attracting  attention  of the physics community for two decades \cite{MS00,BP00,S,McC04}.
During this time a large volume of the empirical researches has been done. Of specific interest is the investigation of the behavior of the market in
 the time periods of the large financial crises when the statistical properties  of the market are drastically distinguished from its properties in quiet days \cite{S,JS98,LM00,S03}. The empirical analysis has  found the occurrence of a number of  peculiarities in market dynamics appearing in these periods. Originally they have been revealed when investigating  the large financial crash  on 19 October 1987 (Black Monday) at New York Stock Exchange. However these  peculiarities  are not specific  for October 1987 and have been later revealed in the financial crashes of 1997 and 1998 years for different countries and Exchanges \cite{S03}.

Firstly, it has been established that the large financial crashes are outliers which are not possible within the scope of the typical price distributions. For their realization the complementary factors absent in quiet days are needed  \cite{JS98,S03}.

Secondary, the emergence of the certain periodic patterns in the dynamics of the different financial assets are detected both before the crash, and immediately after it. In particular, a log-periodic oscillations in time evolution of the price of an asset appear before the extreme event  \cite{S03,SJB96,SJ97,JS99,FF96,VAB99}.  The aftershock period is characterized by an exponentially decaying sinusoidal behavior of a price  \cite{S03,SJB96}. The characteristic behavior of volatility  is also revealed after the crash. In the moment of the crash of October 1987  the implied volatility  of the S\&P500 index makes a shock and then sharply decays as power law decorated with log-periodic  oscillations \cite{S03,SJB96}. In the day of the  maximum drop  the central part  of the empirical return distribution  moves toward negative returns and then begins to oscillate between positive and negative returns \cite{LM00}.

At last, the relaxation dynamics of an aftercrash period is characterized by what is termed  as the Omori law which shows that the rate of of the return
shocks larger than some threshold decays as the  power law with exponent close to 1 for different thresholds  \cite{LM03,WW07}.

There are a large number of works devoted to the modeling of the market dynamics immediately before  crash \cite{S03,SJB96,SJ97,JS99,FF96,VAB99,ML07,GF08} (see
also references in the article of Sornette \cite{S03}). These works based on the empirical observations  exploiting the analogy with critical
 phenomena have been directed to the investigation  of a possibility of predictions of the large financial crash on the base of the modeling of the
 dynamics of  different financial indexes (S\&P500, DJ, etc).

The "microscopic" approach has been suggested in the work \cite{BC98} where the model of the market dynamics taking into account a possibility of occurrence of the crash is developed. In physical terms, the model is reduced to a motion of a Brownian particle in the stationary cubic potential and crashes are considered as rare activated events.

In Refs. \cite{BVS06,BVS07} the generalization of the above model has been suggested for the case of the stochastic volatility which was considered within the scope of the modified Heston model. Physically, the model represents an overdamped Brownian particle moving in the stationary cubic potential under the influence of the fluctuating noise intensity.

Since the statistic properties of the volatility in the Heston model are uniform in a time \cite{DY02},  both  the above models describe  the uniform stochastic process.  In financial terms,  this means  that statistic properties of the market are the same  regardless of where  market is either  in normal regime or close to a crash.  It seems this is poorly consistent  with the assertions that crashes are outliers and the statistic properties of the market are drastically change in extreme days \cite{JS98,S03}. While within the scope of Brownian  motion the assertion that crashes are outliers and  strong nonstationarity  of the market revealed in extreme days \cite{LM00} lead to the conclusion that an "external" time-dependent potential  perturbing  the stationary potential profile, in which a fictitious Brownian particle moves in the case of typical days, is triggered in the vicinity of the crash.  It is likely the appearance of such potential could be traced developing the microscopic approach
in the spirit of the work \cite{BC98}. It is worth  saying  that the authors of this work have  also noted that in the vicinity of a crash
 additional market mechanisms , which have not been  considered in \cite{BC98}, must be taken into account to remove divergences emerging in their model of the aftershock dynamics.

 Another way to detect the influence of the "external" nonstationary perturbation is to analyze the empirical data concerning  the extreme events. It is such approach that will be considered in the given work for the modeling of the market dynamics during the large  financial crash of October 1987.

 The paper is organized as follows. In Sec.\ref{sec:Mod} in terms of the stochastic differential equation (SDE) Ito we consider the model describing the market dynamics during  a crash. Sec.\ref{sec:Estimat} is devoted to the analysis of the empirical data and the estimation of the Kramers-Moyal coefficients. In Sec.\ref{sec:Dis} the discussion of results and conclusion are given.

\section{\label{sec:Mod}The Model}

We consider the model in which immediately before a crash and in the short aftercrash period the dynamics of a financial assets is described by the SDE Ito:
\begin{eqnarray}
&&dx = D_1(x,t)dt + \sqrt{D_2(x, t)}\; dW(t) , \label{eq:eq1}\\
&&D_1(x, t) = - \frac{\partial U(x, t)}{\partial x}\, ,
\label{eq:eq2}
\end{eqnarray}
where $x(t) = \ln S(t + 1)/S(t)$ is the daily log-return, $S(t)$ is  the price of the asset at time $t$, $W(t)$ is a standard Wiener process,
$D_{1}$ and $D_{2}$ are the drift coefficient and the diffusion coefficient, respectively. Physically, Eq.~(\ref{eq:eq1})
describes a motion of an overdamped Brownian particle in a potential $U(x,t)$
\cite{G97}, introduced by Eq.~(\ref{eq:eq2}),with a variable noise intensity depending on a particle position. Like the specific  models  considered
in  \cite{BC98,BVS06,BVS07} Eq.~(\ref{eq:eq1}) presents a Markovian process.

As it has been noted above in the moment of the crash  of October 1987 the implied volatility of the S\&P500 index makes a shock  with  subsequent relaxation to its beforeshock level during several weeks. Simultaneously after the index decline, its "stabilization" is approximately reached within the same time  \cite{S03,SJB96}. All this allows to consider  that a peculiar  synchronization of their movements  occurs during the crash.  In turn, this circumstance makes it possible to assume that in this period  the asset volatility can be considered  as a function of the return $x$
and employ the one-dimensional model ~(\ref{eq:eq1}) with the variable   diffusion coefficient $D_2(x,t)$ being a measure of the volatility.

As it is known   \cite{G97} the coefficients $D_1$ and $D_2$ from SDE ~(\ref{eq:eq1}), defining the the stochastic process $x(t)$,coincide  with the Kramers-Moyal coefficients  and are given by the equations
\begin{eqnarray}
&&D_k(x, t) = \lim_{\tau\rightarrow 0} {\frac{1}{\tau}M_k(x, t, \tau) }\quad (k=1, 2) \label{eq:eq3}\\
&& M_k (x, t, \tau)= \int dx'(x' - x)^k P(x', t + \tau | x, t),
\label{eq:eq4}
\end{eqnarray}
where $P(x', t + \tau | x, t)$ is the conditional probability distribution function of $x(t)$.

Originally, within  a financial context, the method of the estimation of the Kramers-Moyal coefficients directly  from the empirical data has been given in Refs. \cite{FPR00, RPF01}.Further this method has been applied to different financial assets and exchange rates \cite{AI03,BC05,NP07,Gh07,F07,CRA07,LK08}. In all cases the underlying stochastic process is supposed to be stationary (uniform) and the reconstruction the drift  coefficient $D_1$ from the empirical data gave the linear dependence on corresponding return. According to Eq. ~(\ref{eq:eq2}) this leads to the parabolic potential for a fictitious particle that is typical for the quiet market periods. Here we shall consider the estimation of the coefficients $D_k $ for the case of the extreme events of October 1987 when the market  was in a substantially nonstationary phase \cite{LM00}.
\begin{figure}
\includegraphics{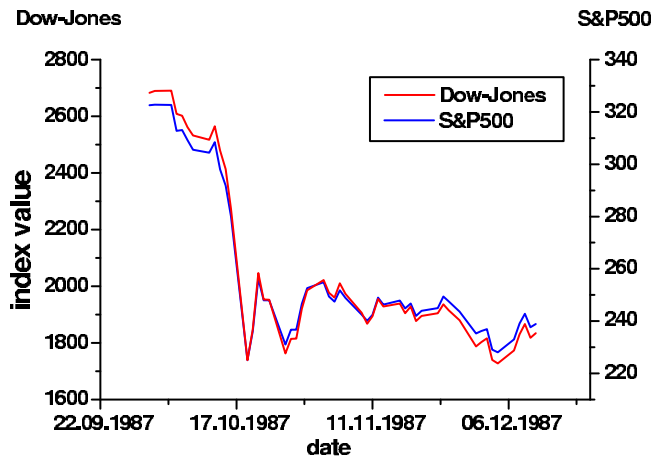}
\caption{\label{fig:fig1} The time evolution DJ and NASDAQ indexes.}
\end{figure}
\begin{figure*}
\includegraphics{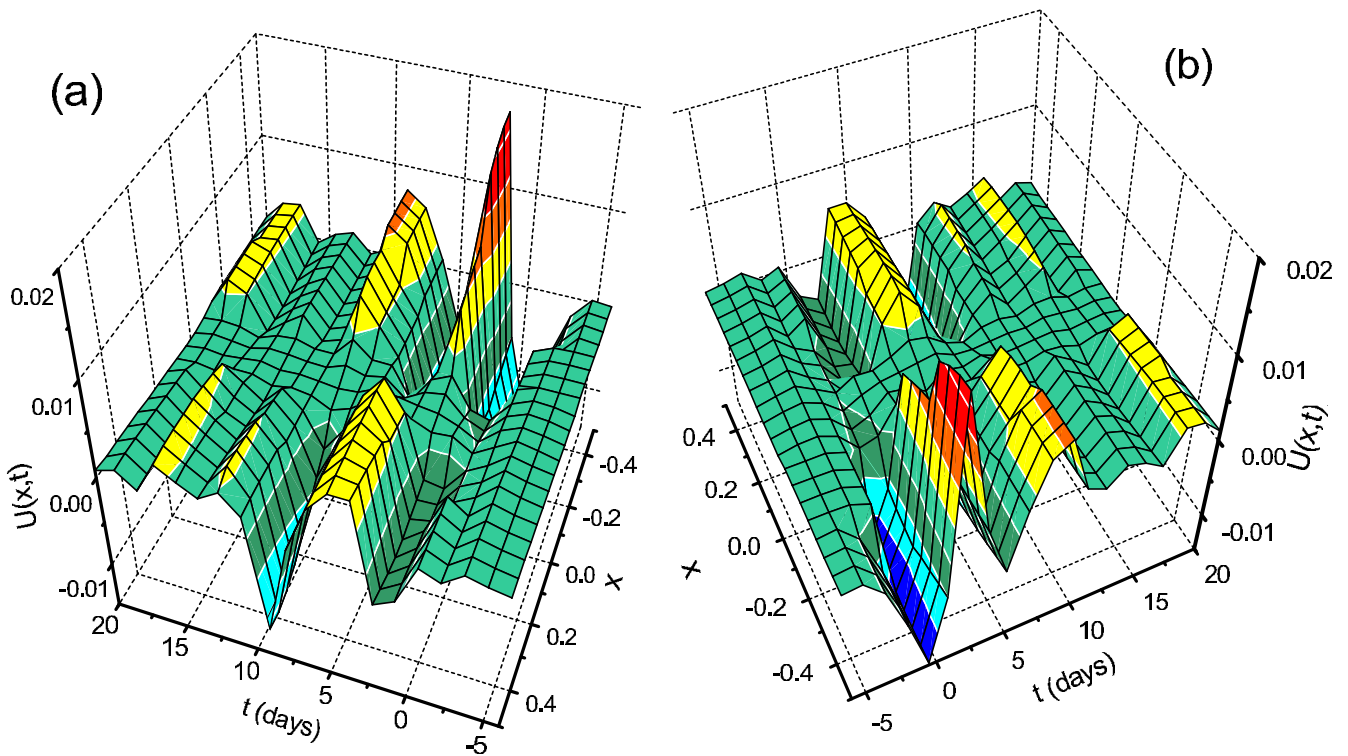}
\caption{\label{fig:fig2} The reconstruction of the potential $U$. In the figures (a) and (b) the potential surface is viewed from two mutually opposite sides}
\end{figure*}

\section{\label{sec:Estimat}The Estimation of  Kramers-Moyal Coefficients}

"\textit{From the opening on October 14, 1987 through the market close on October 19, major indexes of market valuation in the United States declined by 30 percent or more.  Furthermore, all  major world markets  declined  substantially in the month, which is itself an exceptional  fact that contrasts  with the usual modest correlations of returns across countries and the fact that stock markets around the world are amazingly diverse in their organization }" \cite{S03,SJB96}.  In Fig.~\ref{fig:fig1}   the time evolution of two  financial  indexes DJ
and NASDAQ characterizing different sectors of USA economics are
given. Practically   identical dynamics of these indexes during
the crash is seen from Fig.~\ref{fig:fig1}
\begin{figure}
\includegraphics{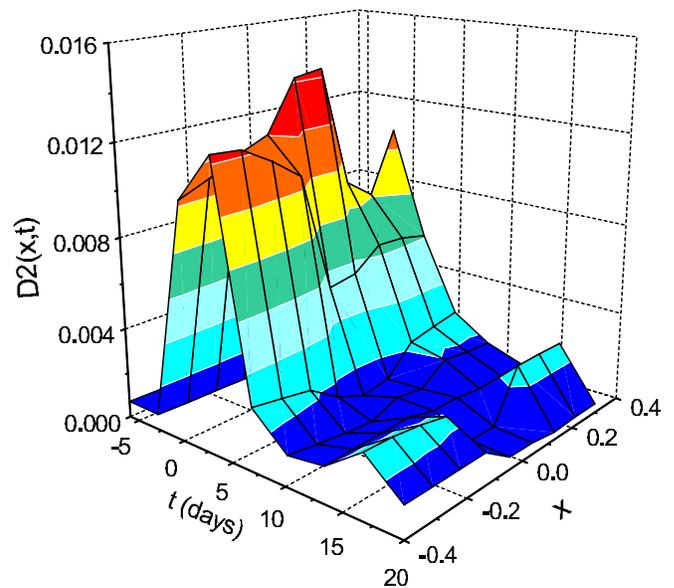}
\caption{\label{fig:fig3}The reconstruction of the diffusion coefficient $D_2$.}
\end{figure}
Such behavior is not accidental and
is well consistent with  the assertion that  large financial
crashes are caused by appearance of substantial cooperation in the
behavior of different markets over the world and local self-reinforcing
imitation between traders \cite{S03}.
These two circumstances led
to the substantial synchronization in the dynamics of different
financial assets simultaneously traded both  in the common market and
in different ones, that  Fig.1 shows. These circumstances allow to
suggest also that log-returns of different stocks simultaneously
traded in markets during the crash have the identical statistical
properties and to consider the probability distributions of the whole
ensemble of such stocks. Actually this approach has been  used in
Ref. \cite{LM00}.

In accordance with  the above said we consider the ensemble which
consists of $n = 467$ stocks
traded in Stock Exchanges NYSE,
NASDAQ and AMEX in the period from 13 October to 13 November 1987 \footnote{The data have been taken from http://finance.yahoo.com.}
each characterized by the daily log-returns
\[x_i(t) = \ln\frac{S_i(t + 1)}{S_i(t)}\qquad i = 1,2\ldots n \, ,\]
where $S_i(t)$ is the close price of $i$ - th asset on day $t$
and the time is measured from the moment of the crash of 19 October.

Apart from the assumption that during  the crash the statistic
properties all $x_i(t)$ are identical  we shall introduce a
fictitious index $S^{*}(t)$ and consider the empirical data $x_i(t)$
$(i =  1,2, ...n)$ as the set of "experimental" values of the  log-returns of this
index detected at moment $t$.  Further the estimation of the
Kramers-Moyal coefficients will be  performed just for the
index $S^{*}(t)$ with log-return $x(t)$.

According to Eq.~(\ref{eq:eq3}) the calculation  of the coefficients
$D_k$ includes the operation of taking the limit $\tau\rightarrow
0$. For the case of the high-frequency sampling a sufficiently exact estimation can be obtained
merely as ratio
\begin{equation}
D_k(x, t) \simeq\frac{M_k(x, t; \tau)}{\tau}\hspace{1cm}
(\tau\rightarrow 0)\label{eq:eq5}
\end{equation}
On the long time scale with the minimal step $\tau = 1$  day, Eq.~(\ref{eq:eq5})
can be broken down. Nonetheless, the rough but, as it will be seen later, giving definite  information
 estimate of the coefficients $D_k$ can be also obtained in this case at  $\tau = 1$. Actually the
approximation is used, meaning an extension of the linear dependence on the long time scale:
\begin{equation}
M_k(x,t; \tau ) = \tau D_k(x, t)\hspace{1cm} (0\leqslant\tau
\lesssim 1 )\label{eq:eq6}
\end{equation}

\begin{figure}
\includegraphics{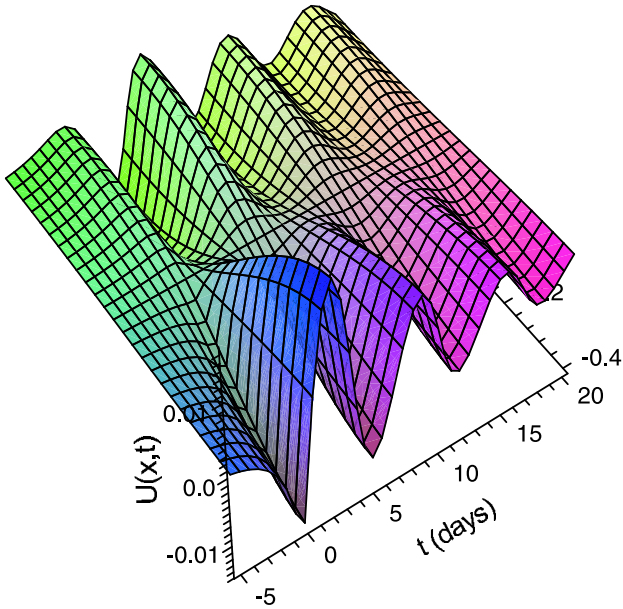}
\caption{\label{fig:fig4}The potential surface $U$ given by Eq.~(\ref{eq:eq7})}
\end{figure}
Thus the estimate of the Kramers-Moyal coefficients has been taken from equality $D_k(x,t) = M_k(x,t,; \tau = 1)$.
For the calculation of the conditional moments $M_k$ the  empirical conditional densities
$P(x_2 t_2|x_1 t_1) = P(x_2 t_2,  x_1 t_1)/ P(x_1 t_1)$ $(t_2 >t_1)$ have been first found where
$P(x_2 t_2,  x_1 t_1)$ and $P(x t)$  are the joint two-point probability density and the one-point probability density, respectively. Then the numerical integration has been performed  in  Eq.~(\ref{eq:eq4}). Once the coefficient $D_1$ has been found the potential $U$ is determined
from Eq.~(\ref{eq:eq2}). The results of the calculations of $U$ and $D_2$ are shown in Figs.~\ref{fig:fig2} and ~\ref{fig:fig3}.

The periodic structure of the potential surface is  clearly distinguishable in  Fig.~\ref{fig:fig2}.  In the regions $x > 0$ and $x < 0$ alternation of wells and hills is observed, the sizes of which  rapidly decrease with time. The periodic structures  of these two regions are displaced in relation to each other so that at the displacement along $x$ axes a well in the range $x > 0$ transfers to a hill in the  range $x < 0$ and conversely.  At $t = 0$ there is  the well of the largest  depth in region $x < 0$ corresponding to the moment of the crash.
The behavior of the potential surface with indicated features reconstructed from the empirical data is well enough   reproduced
 by the smooth surface presented in Fig.~(\ref{eq:eq4}) and given by the equation:
 \begin{widetext}
\begin{eqnarray}
   U(x, t)=  \phi(x, t)
\left\{
     \begin{array}{ll}
         A \sin \omega x \, e^{-\alpha x + \beta_1 t},& t<0  \\
         B\sin\omega x\,\sin(\omega_1 t + b)\, e^{-\alpha x - \beta_2 t},& t>0, \label{eq:eq7}
       \end{array} \right.\\
       \nonumber
   \end{eqnarray}
   \end{widetext}
where
\begin{equation}
 \phi(x, t) = 1 - a\, x^{1/3}\, e^{-\gamma |t|}\label{eq:eq8}
 \end{equation}

The potential $U$ presented by Eq.~(\ref{eq:eq7}) exponentially decays with time and included  trigonometrical functions reproduce
the observed periodic structure. Function $\phi (x, t)$ is largely used for the smooth joint of two surfaces from  Eq.~(\ref{eq:eq7}) at $t = 0$
and rapidly goes to one with the increasing $|t|$.  The parameters Eqs.~(\ref{eq:eq7}) and ~(\ref{eq:eq8}) are determined by comparison with the reconstructed surface on the base of the least-squares method  and presented in  Table~\ref{tab:table1}.

The reconstruction of the diffusion coefficient $D_2(x,t)$ is shown in Fig.~(\ref{eq:eq3}). Qualitatively the behavior of the diffusion coefficient is quite definite,  in spite  of the occurring irregularity. At the moment of the crash   $D_2$ makes a shock with subsequent relaxation
to its beforecrisis  level within a month. In Fig.~(\ref{eq:eq5})  such behavior is reproduced by the  smooth surface with the equation
\begin{eqnarray}
   D_2(x, t)=  \left\{
     \begin{array}{ll}
         A\, t^{- p},& t>0, \label{eq:eq9} \\
         B,& t<0\, ,
       \end{array} \right.
   \end{eqnarray}
 As before the parameters of Eq.~(\ref{eq:eq9}) have been defined by the least-squares method and given in Table  ~\ref{tab:table2}.
 
\section{\label{sec:Dis}Discussion and Conclusion}

\begin{table}
\caption{\label{tab:table1}Parameters of Eq.7 }
\begin{ruledtabular}
\begin{tabular}{cccccccccc}
$A$&$B$& $\omega $&$\omega_1$&$\alpha $&$\beta_1 $&$\beta_2$&$a$&$b$&$\gamma$\\
\hline
$8.6\cdot10^{-3}$&$1.4\cdot10^{-2}$&\mbox{3.9}&\mbox{1.0}&\mbox{0.6}&\mbox{0.5}&\mbox{0.1}&\mbox{0.96}&\mbox{2.5}&\mbox{0.9}\\
\end{tabular}
\end{ruledtabular}
\end{table}
\begin{table}
\caption{\label{tab:table2}Parameters of  Eq.~(\ref{eq:eq9})}
\begin{ruledtabular}
\begin{tabular}{ccc}
$A$&$B$&$ p$\\
\hline
\mbox{$7.6\cdot10^{-3}$}&\mbox{$9.3\cdot10^{-4}$}&\mbox{0.57}\\
\end{tabular}
\end{ruledtabular}
\end{table}
In previous section we have estimated the Kramers-Moyal coefficients $D_1$ and $D_2$  from the empirical data. In spite of the sufficiently rough estimate,the reconstructed  surfaces for the potential  $U(x,t)$ and the diffusion coefficient $D_2$, presented in Figs.~\ref{fig:fig2} and ~\ref{fig:fig3}, can be quite definitely approximated by the smooth surfaces with the needed properties.

Firstly, it has been noted in Introduction that in the aftercrash period the center of the empirical return distribution oscillates
between positive and negative returns \cite{LM00}. Evidently  it is the periodic structure of the potential surface presented in Fig.~\ref{fig:fig4} that provides such behavior.

Secondly, on the base of Eqs.~(\ref{eq:eq1}) and~(\ref{eq:eq2}) with $U$ and $D_2$ from Eqs.~(\ref{eq:eq7}) and~(\ref{eq:eq9}) the time series of the index $S^{*}(t)$ has been generated.
Its  time evolution is presented Fig.~\ref{fig:fig6} by the solid line. It has been obtained by averaging over $150$  the simulated trajectories 
yielding the initial index decline greater than 25\% . As it has been noted in the early observations \cite{SJB96} the aftercrash time behavior of the S\&p500 index is characterized by an expotentially decaing sinusoidal function.  In our case the generated data fit well enough with the function 
\begin{equation}
S(t) = (A_1\,e^{- \alpha_1 t} + A_2\,e^{- \alpha_2 t})\sin (\omega t + \gamma) + A_0\, ,
\label{eq:eq10}
\end{equation} 
incorporating the superposition of two the exponential functions with  the different exponent  $\alpha_1$ and $\alpha_2$.  (the dashed line  in Fig.~\ref{fig:fig6}). The parameters of Eq.~(\ref{eq:eq10}) have been defined by the least-squares method and given in Table  ~\ref{tab:table3}.
  \begin{figure}
  \includegraphics{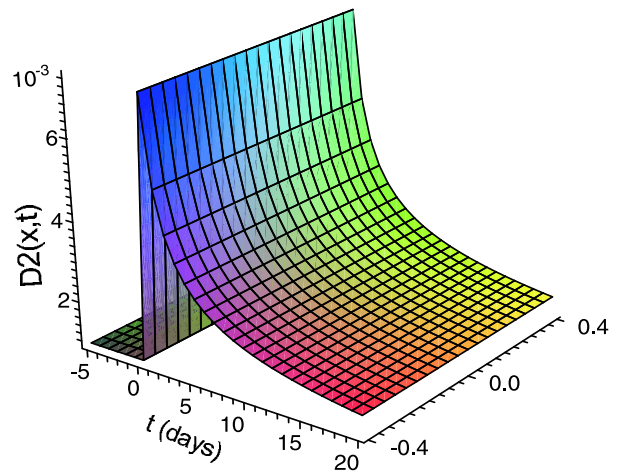}
  \caption{\label{fig:fig5} The behavior of the diffusion coefficient  $D_2$  given by Eq.~(\ref{eq:eq9}).}
  \end{figure}
 
 Thirdly, as  follows from Eq.(\ref{eq:eq9}) at the moment of the crash volatility $ \sim\sqrt{D_2}$  makes a shock and further decays to the aftercrisis level as  power law with the exponent  approximately equal to 0.3. The similar behavior has been also detected for the implied volatility of the S\&P500 index (with exponent   approximately equal to 1.) \cite{SJB96}.
\begin{figure}
\includegraphics{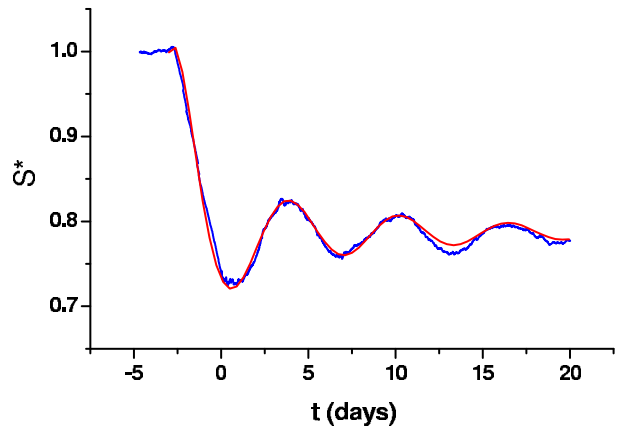}
\caption{\label{fig:fig6} The time evolution of the fictitious index $S^*(t)$. The solid line has been obtained by averaging over 150 the simulated trajectories yielding the initial index decline greater than 25\% . The fit with Eq.(\ref{eq:eq10}) is shown by the dashed line. }
\end{figure}
  
 \begin{table}
\caption{\label{tab:table3}Parameters of  Eq.~(\ref{eq:eq10})}
\begin{ruledtabular}
\begin{tabular}{ccccccc}
$A_0$& $A_1$ &$A_2$ &$\alpha_1$ &$\alpha_2$ &$\omega$ &$\gamma$\\
\hline
\mbox{0.787}&\mbox{0,05}&\mbox{0.031}&\mbox{0,09}&\mbox{0,64}&\mbox{1}&\mbox{- 2.41}\\
\end{tabular}
\end{ruledtabular}
\end{table}
 
 \begin{figure}
\includegraphics{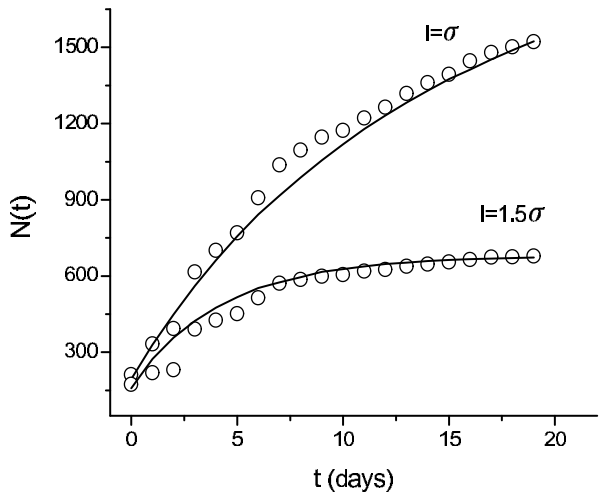}
\caption{\label{fig:fig7} Cumulative number $N(t)$ of log-returns in interval $[0, t]$, whose absolute values exceed a given threshold value $\ell$ for  $\ell = \sigma $ and $1.5\sigma$, where $\sigma = 0.0685$ is standard deviation of log-return $x(t)$  computed over the entire period of 25 days. The generated data  are shown by circles, the solid lines are power law fits.  }
\end{figure} 

Finally, we have examined the fulfillment of the Omori law  within the framework of the given model. To this end the cumulative number  $N(t)$
 of the log-returns in interval $[0, t]$, whose  absolute values exceed a given threshold value $\ell$ has been calculated  for different $t$.
As it is seen from Fig.~\ref{fig:fig7}  the generated data are
described well enough by power law ${t^{1 - \Omega}}$, that is
characteristic for the Omori law, with $\Omega = 0.626$ for the
threshold $\ell = \sigma$ and $\Omega = 0.704$ for $\ell =
1.5\sigma$  where $\sigma = 0.0685$ is the standard deviation of
log-return $x(t)$  computed over the entire period of 25 days. The values of $\Omega$ are well
consistent with the data found by Weber at al \cite{WW07}.
 
 Thus  the presented model, it seems, allows to describe the best known  empirical facts relative to the aftercrash dynamics. To some extent this circumstance can serve as validation of performed estimation of the Kramers-Moyal coefficients.
 
 In conclusion, the given work originates  from the empirical observation  that the large financial crashes are outliers and during the
 crash the statistic properties of a market are drastically changed in comparison with its typical properties. Based on these observations
the physical model presents an overdamped Brownian particle with the variable noise intensity defined by the diffusion coefficient $D_2$,
 moving in the nonstationary potential $U$. The market mechanism that triggers such potential is related to an  imitation between traders
increased up to a critical the point  and a panic reigning in a market in the days of the crash. On the base of the empirical data  within the scope of the Markowian approximation the Kramers-Moyal coefficients have been estimated with the subsequent  determination  of the potential  $U(x,t)$. As it has been shown the    given model  reproduces  well enough the best known  empirical observations detected in the days of the large financial crash of October 1987.

\end{document}